%% file: main.tex
\documentclass[conference]{IEEEtran}
\IEEEoverridecommandlockouts

\usepackage{cite}
\usepackage{amsmath,amssymb,amsfonts}
\usepackage{algorithmic}
\usepackage[ruled,vlined]{algorithm2e}
\usepackage{booktabs}
\usepackage{graphicx}
\usepackage{textcomp}
\usepackage{xcolor}
\usepackage{soul}
\usepackage{multicol}
\usepackage{tabularx,booktabs}
\usepackage{hyperref}
\usepackage{stackengine}

\usepackage{array,multirow,graphicx}
\usepackage[T1]{fontenc}
\usepackage{rotating}
\usepackage{nicematrix}
\usepackage{caption}

\usepackage{subcaption}

\newcolumntype{C}{>{\centering\arraybackslash}X} 
\usepackage{lipsum}
\def\BibTeX{{\rm B\kern-.05em{\sc i\kern-.025em b}\kern-.08em
    T\kern-.1667em\lower.7ex\hbox{E}\kern-.125emX}}

\makeatletter
\newcommand{\linebreakand}{%
  \end{@IEEEauthorhalign}
  \hfill\mbox{}\par
  \mbox{}\hfill\begin{@IEEEauthorhalign}
}
\makeatother

\begin{document}



\input{sections/Abstract.tex}

\input{sections/Introduction.tex}

\input{sections/RelatedWork.tex}
\input{sections/Exp_Overview.tex}

\input{sections/Setup_Components.tex}
\input{sections/MachineLearning.tex}

\input{sections/Evaluation.tex}

\input{sections/Conclusion.tex}

\renewcommand\refname{References}
\bibliography{references}

\bibliographystyle{IEEEtran}


\end{document}

%% file: sections/Abstract.tex


\title{ Beam Profiling and Beamforming Modeling for mmWave NextG Networks}



\author{\IEEEauthorblockN{Efat Samir Fathalla \IEEEauthorrefmark{1},
Sahar Zargarzadeh\IEEEauthorrefmark{1},
Chunsheng Xin \IEEEauthorrefmark{1}, 
Hongyi Wu \IEEEauthorrefmark{2}, 
Peng Jiang \IEEEauthorrefmark{1},\\
Joao F. Santos \IEEEauthorrefmark{3},
Jacek Kibilda \IEEEauthorrefmark{3}, and
Aloizio Pereira da Silva\IEEEauthorrefmark{3}}
\IEEEauthorblockA{\IEEEauthorrefmark{1}Dept. Electrical Computer Engineering,
Old Dominion University, Norfolk, VA, 23529, USA}
\IEEEauthorblockA{\IEEEauthorrefmark{2}Dept. Electrical Computer Engineering, University of Arizona, Tucson, AZ 85721, USA}
\IEEEauthorblockA{\IEEEauthorrefmark{3}Commonwealth Cyber Initiative, Virginia Tech, Arlington, VA 22203, USA}

}

\maketitle

\begin{abstract}
This paper presents an experimental study on mmWave beam profiling on a mmWave testbed, and develops a machine learning model for beamforming based on the experiment data. The datasets we have obtained from the beam profiling and the machine learning model for beamforming are valuable for a broad set of network design problems, such as network topology optimization,  user equipment association, power allocation, and beam scheduling, in complex and dynamic mmWave networks. We have used two commercial-grade mmWave testbeds with operational frequencies on the 27 Ghz and 71 GHz, respectively, for beam profiling. The obtained datasets were used to train the machine learning model to estimate the received downlink signal power, and data rate at the receivers (user equipment with different geographical locations in the range of a  transmitter (base station). The results have showed high prediction accuracy with low mean square error (loss), indicating the model's ability to estimate the received signal power or data rate at each individual receiver covered by a beam. The dataset and the machine learning based beamforming model  can assist researchers in optimizing various network design problems  for mmWave networks.


\end{abstract}
\begin{IEEEkeywords}
mmWave, Beamforming, machine learning, 5G and beyond, wireless communication
\end{IEEEkeywords}
\IEEEpeerreviewmaketitle

%% file: sections/Introduction.tex
\section{Introduction} \label{introduction}
\IEEEPARstart{T}{he} use of high-frequency bands such as the Millimeter wave (mmWave) band has been introduced with 5G and other next generation wireless communication systems to meet the high capacity and throughput requirements ~\cite{lagen2019new}. However, mmWave and higher frequency bands face propagation impairments such as high path loss, diffraction and blockage, rain attenuation, atmospheric absorption, and foliage loss behaviors ~\cite{akyildiz2018combating}. 

To mitigate the aformentioned issues, beamforming technology has been developed to enable the multiplexing of mmWave antennas to provide high spatial processing gains compensating for the isotropic path loss and the other mmWave propagation related challenges  ~\cite{kutty2015beamforming}. Despite being a promising solution, beamforming technology introduces drawbacks related to the high directivity of the formed signals, such as difficulties in random access mechanism and topology management, deciding the number of beams and selecting the best beam directions to cover the user equipments, and the issue of interference from other beams and other base stations (BSs)~\cite{lagen2019new}. Consequently, joint beamforming optimizations among BSs are required to optimize topology  and improve network resource utilization ~\cite{paul2019beamforming,Paul:TMC21}. 

In the literature, analytical approaches have been extensively used to model mobile and wireless communications \cite{7070688,6497025,4723325,6168199,ICC14,6583602}. In this paper, we take a data-driven approach to model mmWave communications using the artificial intelligence and machine learning (AI/ML) technologies.
Adopting AI/ML technologies have shown great potential in addressing beamforming-related challenges~\cite{liang2019deep,zhang2020deep}.  AI/ML models require accurate training datasets in real world settings to ensure better performance and high prediction accuracy for real-world beamforming. However, limited access to experimental mmWave network testbeds leads most researchers to generate synthetic training data through simulations based on various assumptions ~\cite{polese2019millimetera}. 


To address this challenge,  in this paper, we utilize two commercial-grade mmWave testbeds to design an experimental methodology  for mmWave beam profiling to measure the received signal power and data rate at multiple receivers covered by one beam. Furthermore, based on those experimental datasets, we design a  ML prediction model to forecast the signal strength and data date of mmWave networks given a certain topology layout of beamforming and users distributions.  In this paper, the term "beam profiling" refers to the examination of signal strength and transmission quality within a specific location of the receiving node (RX) in relation to the transmitter's (TX) mmWave  beam direction and orientation. The two  commercial-grade mmWave testbeds were provided by theInterDigital Inc. \cite{InterDigital} and the National Instruments Corp. \cite{NI}.

The generation of beamforming datasets is of paramount importance for the research community, particularly in the development of AI/ML models for managing mmWave network topology. A thorough experiment was designed for this paper to acquire broad datasets by dynamically altering the status of the receiving nodes (RXs) surrounding the mmWave beam formed at the transmitting unit (TX). This approach using datasets generated from experimental mmWave testbeds can considerably reduce uncertainties facing simulation-generated data, and reveal previously unstudied mmWave behaviors which can benefit the AI/ML model during the training phase ~\cite{liang2019deep,zhang2020deep}. 

The real-world beamforming datasets from our study could help exploring intelligent solutions for critical mmWave network design problems, such as beamforming optimization, beam scheduling, topology control, user association, power allocation, etc. Ultimately, the research presented in this paper offers valuable insights into the behavior of mmWave networks, which could lead to the development of more effective network management strategies. The main contributions of this study can be summarized as follows:
\begin{itemize}
 \item We developed a comprehensive experimental methodology for creating a mmWave map charting and beam profiling using commercial-grade mmWave network testbeds, an InterDigital 27 GHz mmWave testbed and a 71 GHz mmWave testbed from National Instruments. 
 
 \item In accordance with the experimental testbed, we conducted measurements of the communication quality and signal strength between the transmitting and receiving nodes using a downlink communication model. This enabled us to gather data to create two distinct datasets, both of which were obtained from the mmWave testbeds that were employed in our study. It is worth noting that the experimental design was tailored to ensure that the collected data would be suitable for use in further analysis and machine learning modeling.

 \item Designing and training an ML model to predict received signal strength and communication quality using experimentally collected datasets. The ability to accurately predict signal strength and communication quality is crucial for effectively managing mmWave networks. 

\end{itemize}

The rest of this paper is organized as follows. Section II presents the related work. Section III explains the system overview and mmWave network design. Section IV presents the experimental testbed and apparatus. Section V presents the designed machine learning (ML) model used to predict signal strength and communication quality of  users in the network. Section VI shows the experiment results. Finally, we conclude this paper in Section VII.

%% file: sections/RelatedWork.tex
\section{Related Work}
\begin{table*}
\caption{Experimental testbeds in mmWave Communication Research.}
\label{table2}
\begin{tabularx}{\textwidth}{@{}c*{10}{C}c@{}}
\toprule
\textbf{Ref. and Objective} & \textbf{Frequency (GHz)} & \textbf{Beamforming Generation} & \textbf{Experimental testbed Devices} & \textbf{ML/AI} \\
\midrule

Predict mobility in simple blockage-free networks ~\cite{liu2020learning} & 28.5 & Digital & NI & Yes\\

Investigate beam alignment between TX and RX ~\cite{salehi2020machine} & 60 & Analog  & NI & Yes \\

Investigate joint communications and radar sensing ~\cite{pham2021joint}& 73.5 & Hybrid & NI & No\\

Study and evaluate beam management and tracking ~\cite{jain2020mmobile} & 28 & Analog & Custom testbed & No  \\

Beamforming training approach using MAC layer info ~\cite{shen2021design} & 60 & Analog & Custom testbed & Yes\\

Introduce programmable mmWave testbed ~\cite{joaoSTAMINA2023} & 28 & Digital & InterDigital & No\\

This work studies mmWave TX's beam profiling and propagation behaviour & 27, 71 & Digital & InterDigital, NI & Yes\\

\multicolumn{5}{c}{$^*$ Custom testbeds consist of a Uniform planar array (UPA) with 64 elements, an RF front-end circuit}\\\multicolumn{5}{c}{ using software-defined radios (SDRs), and Field Programmable Gate Arrays (FPGA) for digital signal processing.} \\
\hline
\end{tabularx}
\end{table*}

In recent years, many research efforts investigated the benefits of using mmWave and higher frequency bands for wireless communication systems. The 3rd Generation Partnership Project (3GPP) has also released specifications related to 5G communications that utilize these underutilized frequencies for ultra-low-latency and ultra-wideband wireless communication services. However, researchers have faced difficulties genuinely exploring the mmWave characteristics and basic propagation features due to the lack of a commercial-grade programmable mmWave testbed for experimentation and a shortage of experimental datasets that can be used for AI/ML training phases or other research studies. Several research studies have attempted to address this problem by simulating the mmWave network environment or providing mathematical formulations supporting the research-claimed theories ~\cite{paul2019beamforming,liu2020user,sun2018learning}. 



A few research efforts proposed experiment-based solutions to the problem of having limited commercial-grade resources for mmWave networks. For example, in ~\cite{pham2021joint}, a 5G mmWave beamforming testbed was used to demonstrate angle detection based on beam steering for radar detection. The experimental testbed relied on a baseband processing system implemented on FPGAs and software-defined radio (SDR) devices. In ~\cite{jain2020mmobile}, mMobile developed a custom 5G-compliant mmWave testbed to study and evaluate beam management and tracking algorithms. The authors generated a mobility dataset showing channel measurement metrics, including synchronization signal block (SSB) and Channel State Information-Reference Signal (CSI-RS). 

In ~\cite{salehi2020machine}, the authors utilized an indoor 60GHz mmWave testbed using a national instrument (NI) to examine the challenges of aligning the transmitter (TX) and receiver (RX) beams. The authors generated a dataset comprising the measured SNR metric to evaluate the link quality between TX and RX at different locations and orientations. A convolutional neural network model has been utilized to rapidly identify the locations of the TX/RX nodes and predict the optimal beam pair. On the other hand, researchers in ~\cite{8746643} applied a pre-training method for a deep neural network to predict the mmWave base station (BS) received power. The authors leveraged 3D model simulations to model the signal propagation and then trained the model using transfer learning schemes. 

Table \ref{table2} shows a detailed comparison between a few of the selected existing works in the literature that conducted experimental research in mmWave communications. Although many research efforts have attempted to address the challenges of mmWave networks, only a few have successfully obtained experimental datasets supporting their proposed models. Despite this, the limited availability of diverse datasets highlights the need to bridge the gap between existing datasets and current challenges in mmWave networks. 

To address this issue, it is crucial to employ AI/ML techniques with caution as using biased or limited datasets for training can lead to biased predictions, overfitting, and limited generalization. Insufficient dataset diversity can hinder model accuracy and performance, making it difficult to accurately classify new data in real-world scenarios. This study aims to address this gap by exploring mmWave network mapping and beam profiling, which can aid in topology control and simplify network management and decision-making. To broaden the scope of our investigation, we leverage two mmWave experimental testbeds operating at different frequencies (approximately 27 GHz and 71 GHz). The findings from this experimental testbed aim to facilitate signal strength and communication quality prediction at the UEs sides using AI/ML models.

%% file: sections/Exp_Overview.tex
\section{mmWave Experimental Overview and Methodology}
This paper aims to introduce an experimental approach for evaluating the propagation characteristics of millimeter-wave (mmWave) networks with a focus on the dynamic movement and orientation changes of UEs around BS. The ultimate goal is to create a prediction model that allows the BS to predict the communication quality of nearby UEs based on the current state of the network topology, which can be utilized to enhance the topology strategy and beamforming transmission criteria.

This section outlines the methodology for conducting an experiment to create a dataset showcasing beam profiling for the formed beams at the BS (TX) side. In this context, "mmWave beam profiling" refers to the characterization and analysis of the directional properties of a transmitted beamformed mmWave signal. It should be noted that the beam profiling of the RX node is beyond the scope of this study.

Throughout the article, the terms "UE" and "RX" are used interchangeably to refer to the mmWave beam's receiving node, while "TX" and "BS" denote the mmWave beamforming node that transmits beams in specific azimuth and elevation directions. The study focuses exclusively on line-of-sight (LOS) communications and solely considers the downlink wireless communication mode.

\subsection{Virtual Map Design}
\begin{figure}[tb]
\centerline{\includegraphics[scale=0.27]{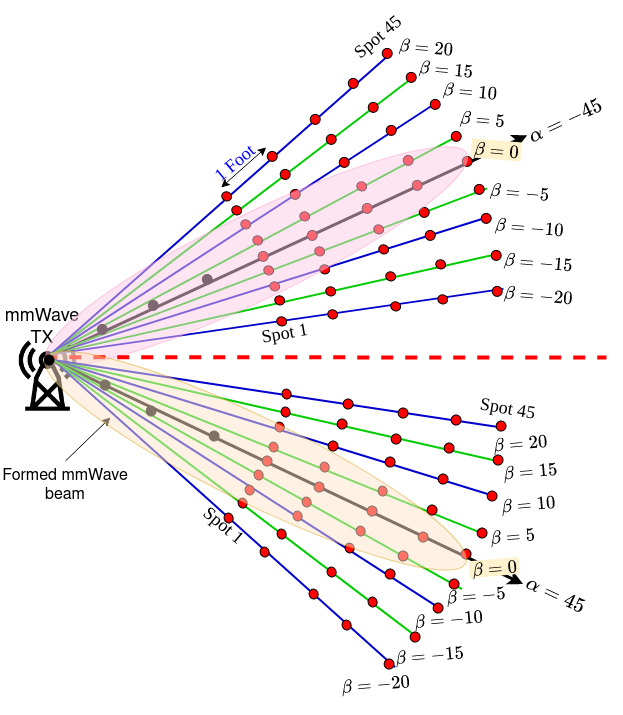}}
\caption{Schematic diagram for the design of mmWave beamforming charting map.}
\label{fig:fig2}
\end{figure}

The study took place in a $10 \times 10$ feet$^2$ unobstructed, open space area where the mmWave TX was located at a fixed corner. The location of the TX was utilized as the reference point for constructing virtual maps around the boresight direction of the individual beams formed at the TX, as depicted in \figurename{ \ref{fig:fig2}}. Each of such virtual maps is composed of various virtual lines positioned around the LOS of the TX's formed beams. These virtual lines are set at different angles, ranging from $-20^{\circ}$ to $20^{\circ}$ with $5^{\circ}$ increments, and contain multiple spots where the RX will be placed. Each virtual line holds a series of identified spots located at one-foot intervals from each other, as shown in \figurename{\ref{fig:fig2}}. 

The experiment involved the use of mmWave TXs that were designed to form beams in various azimuth and elevation directions. For the purposes of simplicity, the elevation direction of the formed mmWave beam was fixed at $0^{\circ}$ throughout the experiment, while the azimuth direction was varied at different angles, as depicted in \figurename{\ref{fig:fig2}}. The RX-generated beam, on the other hand, was fixed in one direction, corresponding to ($0^{\circ}$,$0^{\circ}$) at the azimuth and elevation directions, respectively.

\subsection{Experimental Methodology}
The experiment began by placing the mmWave RX at the first identified spot in the designed virtual map surrounding the formed beam at the TX. The mmWave transmission is then initiated between the TX and RX to capture orthogonal frequency-division multiplexing (OFDM) frames and measure the communication quality strength. After approximately 60 seconds, we stopped the transmission and moved the RX to the next spot, where we repeated the same process. At the end of the experiment, the mean value of the highest measured communication metric obtained at each spot was calculated and appended to its corresponding dataset.

Since the RX would be re-positioned to different measurement locations, the mechanical orientation of the RX would change accordingly. To optimize the exposure of the RX to the mmWave beam formed by the TX at the specified angles and directions, the mechanical orientation of the RX has been adjusted by an angle equivalent to $\alpha \pm \beta$, where $\alpha$ describes the orientation of the formed beam in the azimuth direction and $\beta$ represents the spot angle where the RX was allocated.

During the experiment, we meaured both the received signal reference power (RSRP) and throughput to evaluate communication quality and propagation strength. Throughput measurement reflects the amount of data transmitted and received successfully, which assesses overall performance. In contrast, RSRP measurement quantitatively measures signal strength, a key factor in determining communication quality. Measuring RSRP in different locations and conditions helps identify areas with weak signals or interference, optimize antenna placement, and improve system performance, which ultimately enhances signal quality and coverage and reduces dropped calls.

\subsection{Collected Datasets}
As stated earlier, our experimental study was performed using two distinct mmWave testbeds, namely the InterDigital and NI testbeds. Through the implementation of a virtual map design and the aforementioned experimental methodology, two separate datasets were produced. The first dataset was generated by conducting the experiment with the InterDigital mmWave testbed, where the Received Signal Received Power (RSRP) metric was measured. The second dataset was derived from the NI testbed and illustrates the achieved throughput.

Each data point collected in our study was characterized by several attributes, including $\alpha$, $\beta$, $d$, and the measured value of the chosen communication metric. The parameter $\alpha$ describes the orientation of the formed beam in the azimuth direction, which impacts the map design around the boresight direction of such a beam. The attribute $\beta$ reflects the angle between the boresight direction of the formed beam at one of the transceivers and the location of the fixed formed beam for the other corresponding transceiver, primarily representing the transmitter. This parameter is also critical in determining the optimal beamforming configuration and evaluating the performance of the mmWave system under different conditions. By analyzing the values of $\alpha$ and $\beta$ for each data point, we can determine the optimal beamforming configuration for different scenarios, leading to higher data rates, improved reliability, and better utilization of the available bandwidth.

%% file: sections/Setup_Components.tex
\section{Experiment components and interconnections}
In this section, we explain the adopted testbeds components and interconnections. We first start with explaining the InterDigital testbed components and connections, then explain the design and components of the NI testbed.

\subsection{InterDigital mmWave testbed}

\begin{figure}[tb]
\centerline{\includegraphics[scale=0.3]{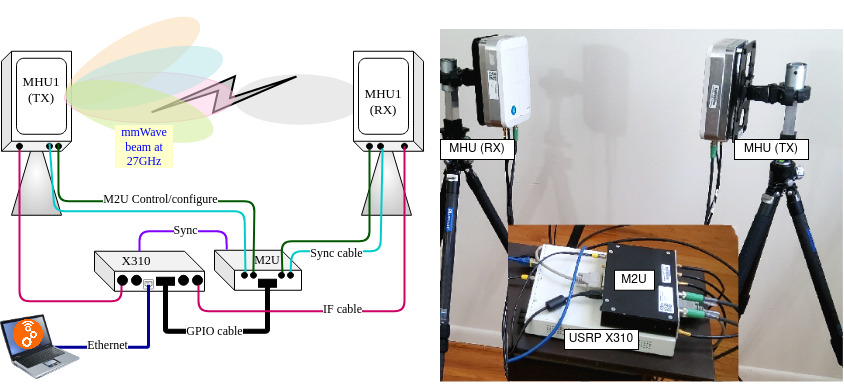}}
\caption{Schematic diagram and actual image for the experiment testbed and apparatus.}
\label{fig:fig3}
\end{figure}

The InterDigital mmWave testbed is composed of a pair of programmable mmWave transceivers known as master head units (MHUs) and a unit controller called the master to USRP (M2U) unit, as shown in \figurename{ \ref{fig:fig3}}. Software-defined radio units, such as Universal Software Radio Peripheral (USRP) model X310 ~\cite{USRP_X310} is used to control and program the InterDigital mmWave device. In summary, each component of the InterDigital mmWave system can be briefly described as follows:

\subsubsection{MHU Head Unit (MHU)} is a versatile component that can function as either a mmWave transmitter or receiver. It is responsible for converting a 5.3GHz intermediate frequency signal to/from the FR2 n257 mmWave spectrum band, which operates around 27 GHz. The MHU's $8 \times 8$ phased array antenna design enables the formation of beams with a range of $\pm45^{\circ}$ in the azimuth and $\pm35^{\circ}$ in the elevation planes, with a resolution of $11.25^{\circ}$ and $11.67^{\circ}$, respectively. In each Master Head Unit (MHU), a pre-generated codebook of 63 beams is stored, where each beam is assigned a unique identifier ranging from beamID 1 to beamID 63. These beams are arranged in a $9 \times 7$ grid. It is important to note that only one beam can be utilized for transmission/reception at any given time.

 \subsubsection{Mast to USRP (M2U)} functions as an intermediary between the MHUs and the software-defined radio (SDR) device, providing power and configuration to the MHUs, including mode of operation (transmit/receive) and selected beam ID configuration. Additionally, the M2U synchronizes the clock signal between the MHUs and the USRP device, and is responsible for translating GPIO commands from the USRP to the MHUs. as,

\subsubsection{USRP and GNU Radio-based framework} 
This study utilized the open-source GNU Radio-based SofTwAre-defined Mmwave INitial Access (STAMINA) framework ~\cite{joaoSTAMINA2023} to manage beamforming and radio transmission in the InterDigital mmWave testbed. The STAMINA framework was employed to configure and control a USRP X310 model, which performed two critical functions: (1) up/down-converting the intermediate frequency (IF) to/from baseband signals, primarily the Orthogonal Frequency Division Multiplexing (OFDM) frames generated by the STAMINA framework, and (2) regulating the voltage levels on the general-purpose input/output (GPIO) interface, which was an essential aspect of overall system control.

In the experimentation with the InterDigital mmWave testbed, the STAMINA framework was used to scan through various beams at the transmitter (TX) with different angles ranging from ($-45^{\circ}$, $0^{\circ}$) to ($45^{\circ}$, $0^{\circ}$) at azimuth and elevation directions, respectively, corresponding to beam ID 28 to 36. In contrast, the receiver (RX)-formed beam was fixed at (0$^{\circ}$, 0$^{\circ}$) in both azimuth and elevation, corresponding to beam ID 32.

\subsection{NI mmWave testbed}
In our study, we employed a second experimental testbed that was provided by the National Instruments (NI) \cite{NI}. The NI testbed depicted in \figurename{ \ref{fig:fig4}} consisted of mmWave transceivers and data acquisition systems (controllers) that facilitated the collection and analysis of experimental data.

\subsubsection{NI Transceivers} are a pair of programmable mmWave transceivers that can operate as either transmitter or receiver. Each NI transceiver can up/down-convert an intermediate frequency (IF) signal ranging from 8.5-13.5 GHz to the mmWave operational frequency range of 71-76 GHz. Further, each NI transceiver can form beams in different directions using a pre-designed codebook consists of 25 beams, covering the azimuth plane from -60$^{\circ}$ to 60$^{\circ}$ with a 5$^{\circ}$ resolution.

\subsubsection{NI Controllers} are responsible for managing the NI transceivers and analyzing the quality of communication. Each controller can configure the transceiver to operate as a TX or an RX and select a particular beam direction from a predefined set of 25 beams spanning the azimuth plane from -60$^{\circ}$ to 60$^{\circ}$ with a 5-degree resolution. 

\begin{figure}[tb]
\centerline{\includegraphics[scale=0.15]{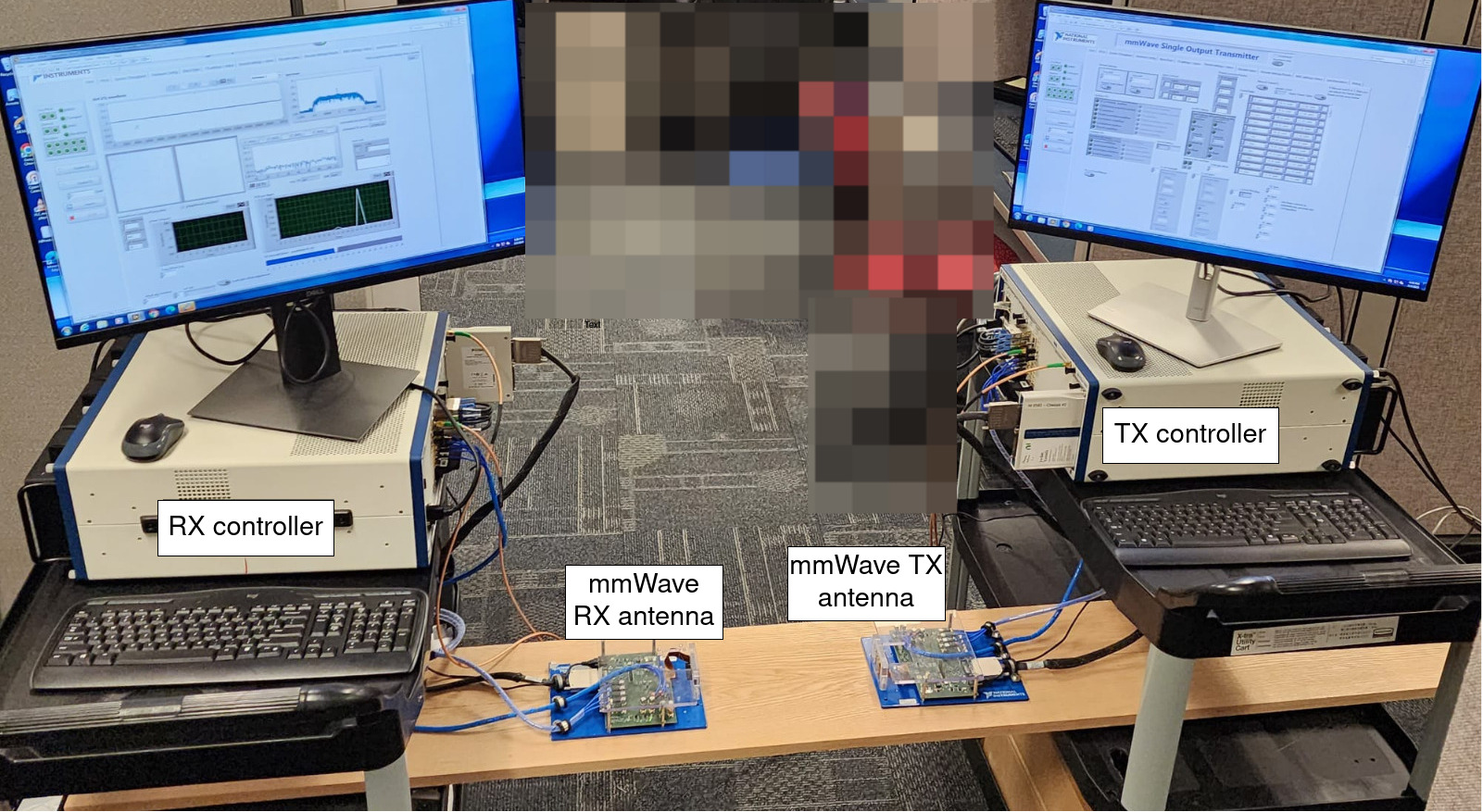}}
\caption{NI experiment testbed. }
\label{fig:fig4}
\end{figure}

The NI experimental testbed and its connections are demonstrated in \figurename{ \ref{fig:fig4}}. The testbed consists of two NI controllers, each directly connected to an NI transceiver that can function as a TX or an RX. The NI mmWave transceiver generates signals using a combination of digital signal processing and analog components. The device's digital signal processor produces baseband signals, which are subsequently upconverted to mmWave frequency range using a frequency synthesizer and mixer. Following this, the amplified upconverted signal is transmitted through the antenna. 

NI developed the control program in LabVIEW that runs on both devices to measure and calibrate IQ impairments for the mmWave transceivers. During the experimental testbed, we sweep through all 25-formed beams available for the TX with different azimuth angles ($\alpha$) ranging from -60$^{\circ}$ to 60$^{\circ}$ with a 5$^{\circ}$ resolution, while keeping elevation angle fixed at 0$^{\circ}$. In the RX experiments, we maintained a fixed location for the TX and varied the position of the Rx based on the angles ($\alpha$+$\beta$). This approach allowed us to study the impact of changing angles on the performance of the Rx, while keeping the TX location constant.

%% file: sections/MachineLearning.tex
\section{Machine Learning Model for Beamforming Modeling}

The main objective of this research is to devise a model that empowers the network's BS to predict the communication quality by leveraging the current status of the UEs position and orientation. This objective can be achieved by integrating a lightweight ML model within the network's BSs to facilitate dynamic network topology redesigns. With the network continuously changing, the ML model can simplify the task of network prediction by enabling adaptive decision-making. For this purpose, a simple yet effective ML model must be selected to deliver accurate predictions under varying network conditions in quick response.

Over the past few years, Neural Network (NN) models have garnered considerable interest due to their adaptability and capacity to grasp intricate non-linear associations between input features and target variables, as evidenced by recent research publications \cite{liu2020user,shen2021design,liu2020learning,kazemi2022channel}. In view of this, we have designed and conducted an exploration of the applicability of NN models to achieve the stated objectives of our investigation; i.e. predicting communication quality strength (RSRP/data rate) given. Datasets obtained from our experimental study using InterDigital and NI mmWave testbeds respectively.

\subsubsection{Overview of the proposed model}
\begin{figure}[ht]
\centerline{\includegraphics[scale=0.22]{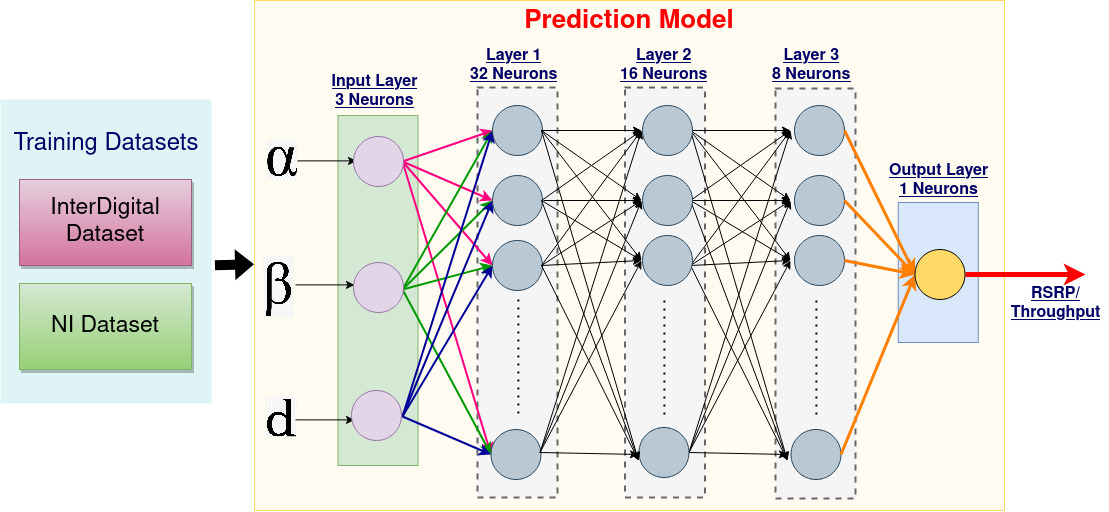}}
\caption{Schematic diagram for the design NN with 3 hidden layers.}
\label{fig:ML_model}
\end{figure}

As shown in ~\figurename{ \ref{fig:ML_model}}, we propose to use the NN model with careful design, which consists of a set of fully connected hidden layers (FCLs) to predict communication quality, instead of using other complicated structures. Simple NN can be trained faster as it has fewer layers and parameters, reducing the computational complexity and providing a faster response to estimate the network throughput and channel quality, which is crucial in wireless communication scenarios. Meanwhile, such an NN network requires careful customization to make sure the model is capable of capturing intricate relationships between our unique inputs and the desired outputs.

Specifically, the first hidden layer in the selected model comprises 32 neurons, with the number of neurons gradually decreasing towards the output layer, which contains a single neuron representing the predicted communication quality strength of the RX. In addition, the structure of the NN, including the number of hidden layers and neurons per layer, should be carefully selected as it is determined by factors such as the number of features and the scale of the dataset. We present an example design with 3 hidden layers in  ~\figurename{ \ref{fig:ML_model}}, in which each hidden layer contains 32, 16, and 8 neurons, respectively. In the next section, we will conduct experiments to identify the optimal NN design for the data collected from InterDigital and NI devices.

\subsubsection{Model input and output}
The input features accepted by the designed NN model include the boresight direction orientation $\alpha$, the RX location's angle $\beta$, and the distance $d$ between the RX and the TX to predict the Reference RSRP or data rate values according to the provided training dataset, as demonstrated in ~\figurename{ \ref{fig:ML_model}}.  It is worth noting that the selection of FCLs is driven by their ability to learn non-linear mappings between input and output data. These layers are effective at capturing intricate relationships between input variables and the desired output, particularly when processing vector inputs. FCL models optimize the weights and biases connected to each layer's neuron. As the input data is pre-processed and reduced to a fixed-length vector representation, FCLs can be applied to one-dimensional data, enabling the layer to map the input vector to the output label ~\cite{ma2017equivalence,malach2020computational}.

\subsubsection{Model activation functions}

We adopt both Tanh and ReLU activation functions for the input and hidden layers of our NN model, respectively. The choice of activation function depends on the specific problem and the characteristics of the data. However, ReLU can model intricate non-linear relationships between inputs and outputs because it is a non-linear activation function. As a result, it serves as a potent tool for input/output ML models that must learn intricate patterns.~\cite{kulathunga2020effects}. 

Due to its steeper gradient, the Tanh activation function can help our NN model achieve faster convergence during training. It is important to note that the vanishing gradient problem, in which the gradient shrinks too much for efficient training, may affect the Tanh function in deep neural networks. However, the Tanh activation function can produce good results for smaller models. The final decision regarding the activation function should be made in light of the details of the problem at hand as well as the characteristics of the data ~\cite{szandala2021review,kamalov2021comparative}. 

In the next section, we evaluate the designed NN model and compare the prediction performance and capabilities of our model with various classical ML models used for predictions.

%% file: sections/Evaluation.tex
\section{Experiment Results and Evaluation}

This section provides a quantitative assessment of the designed mmWave beam profiling experiment presented in this paper. Firstly, we present a concise visualization and overview of selected samples from the obtained datasets obtained during our experimentation with the InterDigital and NI testbeds. Subsequently, we present a comprehensive evaluation of the performance of the designed machine learning (ML) model outlined throughout this investigation, including a detailed comparison with alternative approaches. Table \ref{table1} summarizes the configuration of the presented testbeds. 

\begin{table*}
\caption{mmWave map charting and beam experimental profiling parameters.}
\label{table1}
\begin{tabularx}{\textwidth}{@{}c|*{15}{C}c@{}}
\toprule
\textbf{Metric} & \multicolumn{2}{c}{\textbf{Experiment Testbed}}\\
 & InterDigital  & NI \\ 
\midrule
$\alpha$ range (azimuth,elevation) & ($\pm45^{\circ}$, 0$^{\circ}$)  & ($\pm60^{\circ}$,0$^{\circ}$)\\
$d$ (Feet) & $d=4$ to $d=8$  & $d=1$ to $d=6$\\
$\beta$ range & $\pm20^{\circ}$ & $\pm25^{\circ}$\\
\# spots surrounding boresight direction & 45 spots & 66 spots\\
mmWave operational frequency & 27 GHz & 71GHz\\
IF frequency & 5.3GHz & 8.5 GHz \\
OFDM frames subcarriers &  64 & 64\\
Dataset \# of samples & 405 & 1650\\
Assessment metric & RSRP (dbm)& data rate (Gb/s)\\
\bottomrule
\end{tabularx} 
\end{table*}
\subsection{Dataset Representation of InterDigital mmWave-based testbed}

\begin{figure*}[htbp]
    \centering
    \begin{subfigure}[b]{0.5\linewidth}
\centerline{\includegraphics[scale=0.35]{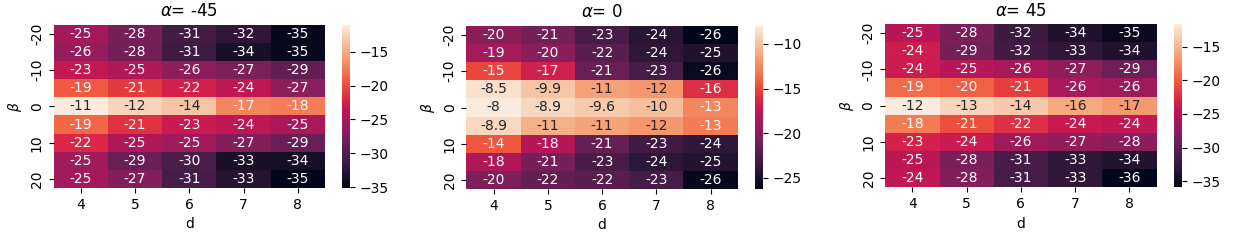}}
        \caption{Heatmap showing RSRP values given the impact of changing $\alpha$, $d$ at different spot angles within the virtual map line for the InterDigital experimental dataset.}
        \label{fig:fig5}
    \end{subfigure}
    \hfill
    \begin{subfigure}[b]{0.5\linewidth}
\centerline{\includegraphics[scale=0.32]{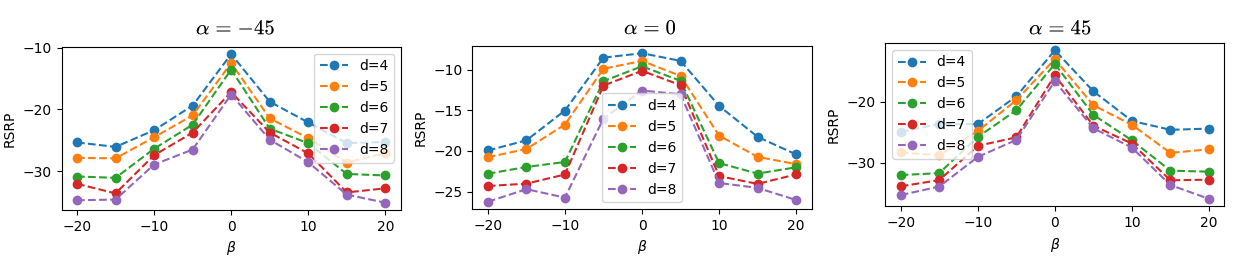}}
        \caption{Formed beam profile at different $\beta$ for the InterDigital-based experimental dataset.}
        \label{fig:fig6}
    \end{subfigure}
    \caption{InterDigital-obtained dataset representation.}
    \label{fig:mainfig}
\end{figure*}

In light of the previous explanation, \figurename{ \ref{fig:mainfig}} analyze samples of the measurements obtained from the first dataset using the InterDigital mmWave testbed. \figurename{ \ref{fig:fig5}} illustrates how the RSRP values vary when the RX is relocated in the designated map spots with different $\beta$ angles across the TX's formed beams with different $\alpha$ values. For demonstration purposes, we selected to present the beam profile of the formed beams at $\alpha=-45^{\circ}$, $\alpha=0^{\circ}$, and $\alpha=45^{\circ}$.

As depicted in \figurename{ \ref{fig:fig5}}, the RSRP value is significantly affected by the RX's orientation and distance from the TX. This behavior can be explained given the relationship between the transmitted and received power, which can be mapped to the RSRP profiling obtained during the experiment. 

Since the interdigital mmWave transceiver uses digital beamforming, the precoding weight ($W$) relies on base-band weights ($W_{BB}$) and steering phase weights ($W_{PS}$) ~\cite{roze2016comparison,kebede2022precoding}. The steering weight ($W_{PS}$) is associated with the steering angle of the generated beam with angle $\alpha$. On the other hand, the base-band weight ($W_{BB}$) is responsible for the amplitudes of the generated beams. Since RSRP is defined as the average power of the reference signal received over the resource elements carrying cell-specific reference signals within the considered measurement bandwidth, then we can derive the RSRP (dB) formula, which can be calculated as:
    \begin{equation}
RSRP= 10\log_{10}\ \frac{1}{N_{ref}} \ P^T \ |(\sqrt{\frac{M}{PL}} \zeta \ W_{BB} \ F_M(\beta)|^2
    \label{eq:4}
    \end{equation}
where $N_{ref}$ is the receiver's thermal noise power spectral density. $M$ is the number of antennae in the InterDigital 8X8 uniform rectangular array, $PL$ is the RX average path loss, $\zeta$ is the complex power gain of small-scale fading between TX and RX modeled as Gaussian distribution, and $F_M(.)$ is the Fege\'r kernel function.


To explain the data presented in \figurename{ \ref{fig:fig5}}, it is important to consider the parameters that impact the received power and the RSRP value. One of these parameters is $PL$, which is impacted by the operational frequency and distance between the transmitter and receiver. Additionally, the channel response is sensitive to the steering angles between the TX and RX and the spot angle of the RC ($\beta$), which affects the effective channel gain and, consequently, the received power. Another factor to consider is the preconfigured values of base-band weights ($W_{BB}$), which is different for each formed beam direction $\alpha$; i.e., it has a noticeable effect on the obtained RSRP values. 

Further analysis showing the detailed beam profiling for the formed beams at various $\alpha$ directions can be seen in \figurename{ \ref{fig:fig6}}. The figure highlights how the RSRP value of the signal at each point in the designed map around the formed beam boresight direction. The RSRP values decrease when the angle difference between the boresight direction of the generated beam and the RX's location ($\alpha-\beta$) increases, showing its maximum values when ($\alpha-\beta$) value is equal to 0.

\subsection{NI mmWave testbed dataset Representation}

\begin{figure*}[htbp]
    \centering
    \begin{subfigure}[b]{0.5\linewidth}
\centerline{\includegraphics[scale=0.33]{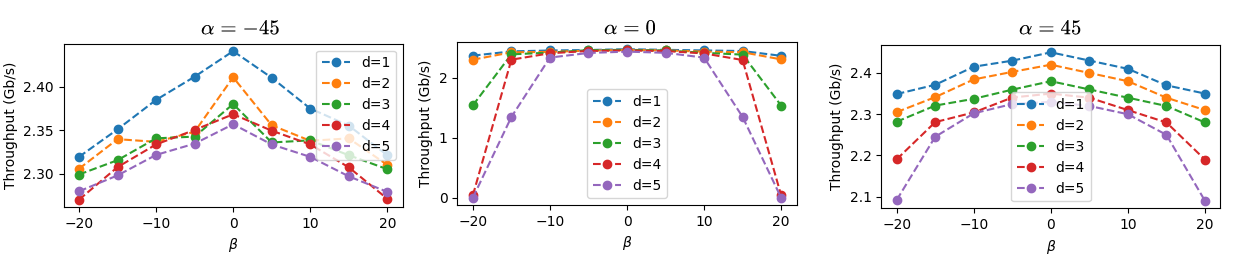}}
\caption{data rate values for NI experiment given the impact of changing $\alpha$, $d$ at different spot angles within the virtual map line.}
\label{fig:fig7}
    \end{subfigure}
    \hfill
    \begin{subfigure}[b]{0.5\linewidth}
\centerline{\includegraphics[scale=0.33]{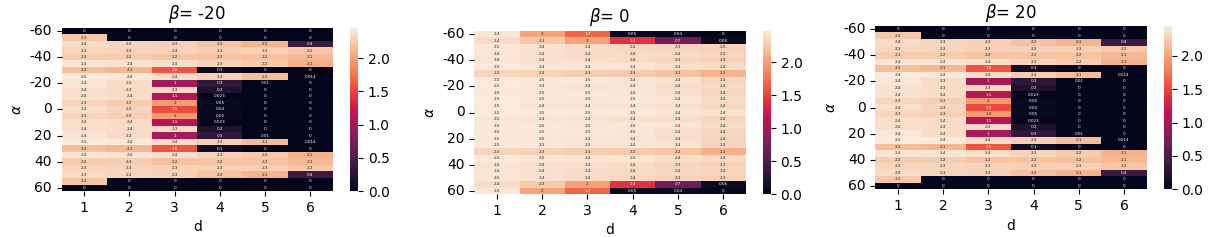}}
\caption{Heatmap showing data rate values for NI experiment given the impact of changing $\alpha$ and spot angle $\beta$.}
\label{fig:fig8}
    \end{subfigure}
    \caption{InterDigital-obtained dataset representation.}
    \label{fig:mainfig2}
\end{figure*}

\figurename{ \ref{fig:mainfig2}} show samples of the measurements representation of the second dataset obtained using the NI mmWave testbed. \figurename{ \ref{fig:fig7}} shows the relationship between the measured data rate when varying the distance $d$ at different $\beta$ angles, while altering RX-generated azimuth direction $\alpha$. Given the representation shown in \figurename{ \ref{fig:fig7}}, the data rate significantly reduces when the distance decreases or the angle $\beta$ where the RX is allocated deviates significantly from the TX-directed beam. For example, when the angle $\beta=0$, the highest data rate values will be obtained. In contrast, when $\beta$ angles increase, the communication data rate will significantly reduce. Further, due to the high operational frequency (70 GHz), the communication data rate between TX and RX depletes until it vanishes completely when the distance between TX and RX exceeds 5 feet.

To better visualize the obtained dataset using the NI-based experimental testbed, \figurename{ \ref{fig:fig8}} shows a detailed heatmap for measured data rate value given the variation of the azimuth direction ($\alpha$) for the RX, the distance ($d$), and selected $\beta$ values. As presented in \figurename{ \ref{fig:fig8}}, the impact of the similar absolute $\beta$ angles is almost the same on the measured data rate value. For example, when $\beta=-20^{\circ}$, the attained data rate is almost the same when $\beta=20^{\circ}$ for the same $\alpha$ direction.

\subsection{Evaluation of the proposed model}

\subsubsection{Overall Performance}
\begin{figure}[h]
\centerline{\includegraphics[scale=0.26]{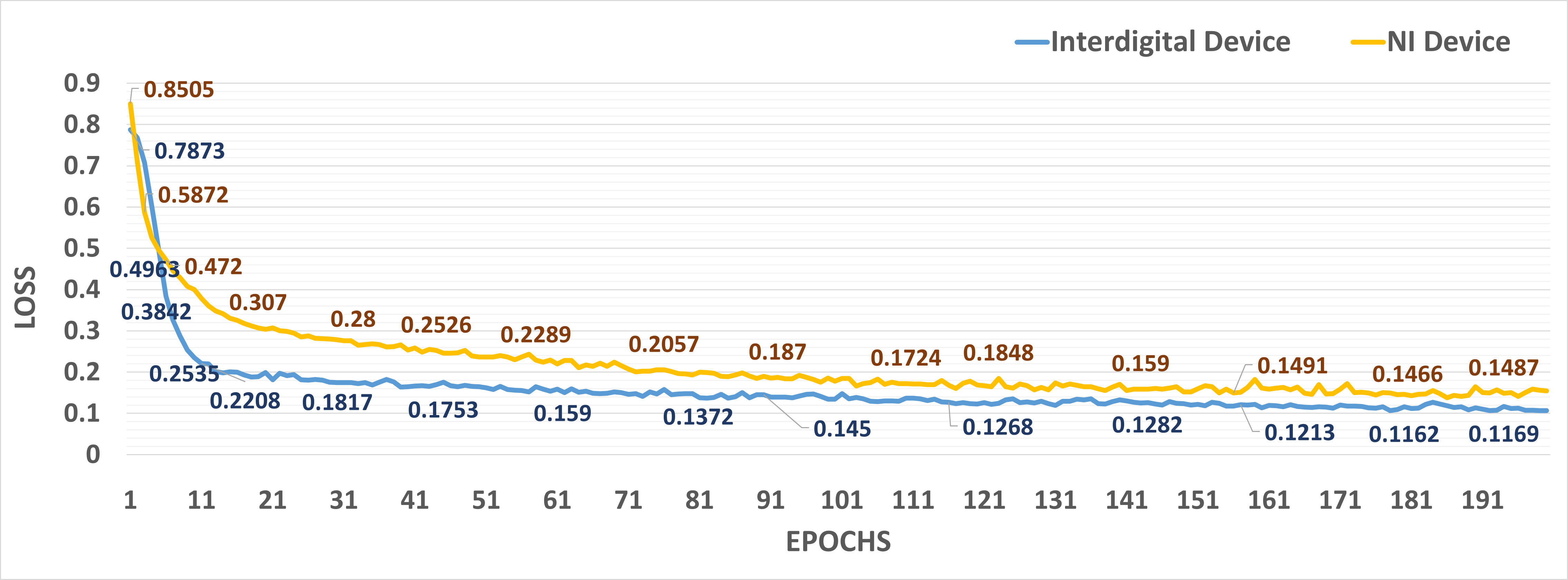}}
\caption{The training process on InterDigital and NI datasets, respectively. }
\label{fig:Loss}
\end{figure}

As explained in the previous section, we design a simple yet powerful NN model to predict the communication quality strength given the experimentally obtained datasets presented in this paper. To train the model, we partitioned the measured datasets into two sets - 80\% for training and the remaining 20\% for testing and validation. The training dataset was used to optimize the weights and biases of the NN model, while the testing dataset was employed to assess the model's generalization performance on unseen data. The designed model was trained for 200 epochs with a batch size of 10 samples. Figure \ref{fig:Loss} plots the training losses on both datasets.  

To evaluate the performance of the designed NN models, two performance metrics, the training Mean Squared Error (MSE) and R-squared (R2), were used ~\cite{sun2018learning}. MSE measures the average squared difference between the predicted and actual values in the testing dataset, providing an absolute measure of error. The InterDigital dataset provides RSRP data, with measurements in units of dBm. Therefore, the MSE is computed based on these dBm values. Similarly, for the NI dataset results, the MSE is computed based on the data rate values of the provided data. In contrast, R2 measures the proportion of the variance in the target variable that can be explained by the model, providing a relative measure of how well the model fits the data. The customized NN model shows a relatively low MSE during the training/testing phases for both InterDigital and NI-obtained datasets, with an average MSE of 0.16 and 0.17, respectively, as depicted in \figurename{\ref{fig:ML-comparison}}.  

\subsubsection{Optimal Structure}
To investigate the performance of adopting various structures in NN models, we compared the performance with various architectures, such as the number of hidden layers and neurons per layer, in the context of a prediction task on two datasets. The results in Table \ref{table_NN_Layers} show the R2 performance of each design. 

From Table \ref{table_NN_Layers}, we observe that the NN with five hidden layers outperformed the other NN architectures, suggesting that a deeper network can learn more intricate representations of the input data, thereby enhancing its accuracy in predicting the outcome of the task. Adopting a NN with limited hidden layers would fail to capture complicated patterns in the communication system, while introducing too many layers (such as 6 hidden layers) may hurt the performance, e.g., due to overfitting.

\begin{table}[h]
\renewcommand{\arraystretch}{1}
\caption{The R2 performance of the designed model with different hidden layers (exclude input and output layer) and neurons on InterDigital and NI datasets.}
\label{table_NN_Layers}
\centering
\begin{tabular}{|Cp{1.8cm}||p{2cm}|Cp{1.5cm}|Cp{1.3cm}|}
\hline
\textbf{\# H. Layers} & \textbf{\# Neurons} & \textbf{InterDigital} & \textbf{NI}\\  
\hline
1 & 64 & 0.81 & 0.84\\
1 & 32 &  0.82 & 0.85 \\
2 &	32,16 & 0.88 & 0.89 \\
3 &	32,16,8 & 0.94 & 0.93 \\
4 & 32,16,8,4 & 0.95 & 0.94 \\
\textbf{5} & 32,16,8,4,2 & \textbf{0.98} &\textbf{0.96} \\
6 & 64,32,16,8,4,2 & 0.95 & 0.93 \\

\hline
\end{tabular}
\end{table}

\subsubsection{Comparison with classical ML models}

Moreover, to ensure the optimality and effectiveness of our designed model, we compared it with various classical ML models commonly employed for prediction purposes. The selected classical ML models used for such a comparison are: Linear regressions, CatBoostRegressor, RandomForestRegressor, and GradientBoostingRegressor Linear regression is a statistical technique used to model the linear relationship between one or more independent variables and a dependent variable. On the other hand, CatBoostRegressor and RandomForestRegressor are ensemble methods that use decision trees to enhance prediction accuracy. These models utilize multiple decision tree outputs to make more accurate predictions. Another ensemble method based on decision trees is the GradientBoostingRegressor, which employs gradient boosting to improve the performance of individual trees.

\begin{figure*}[tb]
     \centering
     \begin{subfigure}[b]{0.45\textwidth}
        \centerline{\includegraphics[scale=0.38]{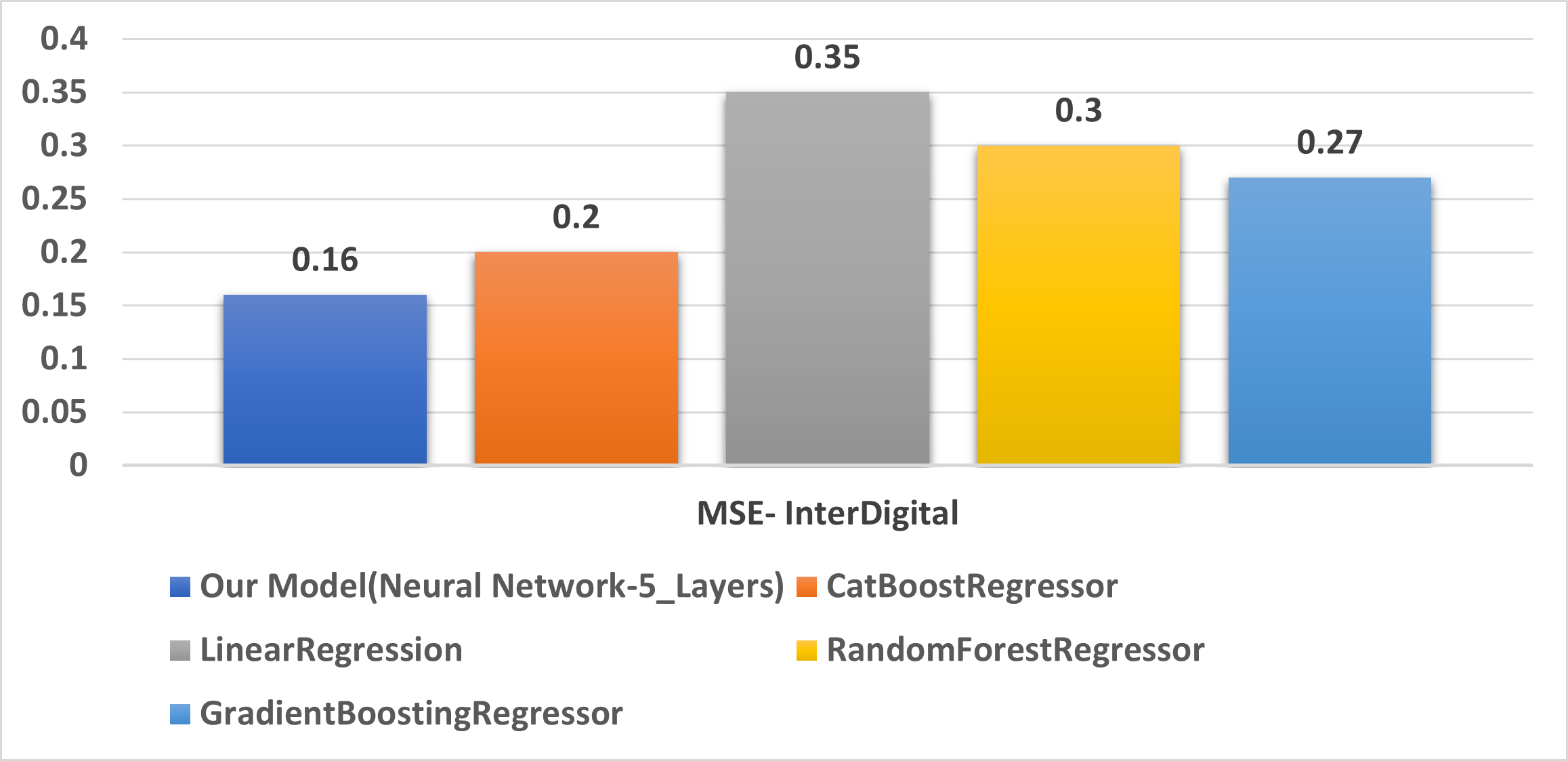}}
        \caption{MSE value for InterDigital dataset trained ML models.}
        \label{fig:MSE-Inter}
     \end{subfigure}
     \hfill
     \begin{subfigure}[b]{0.45\textwidth}
        \centerline{\includegraphics[scale=0.4]{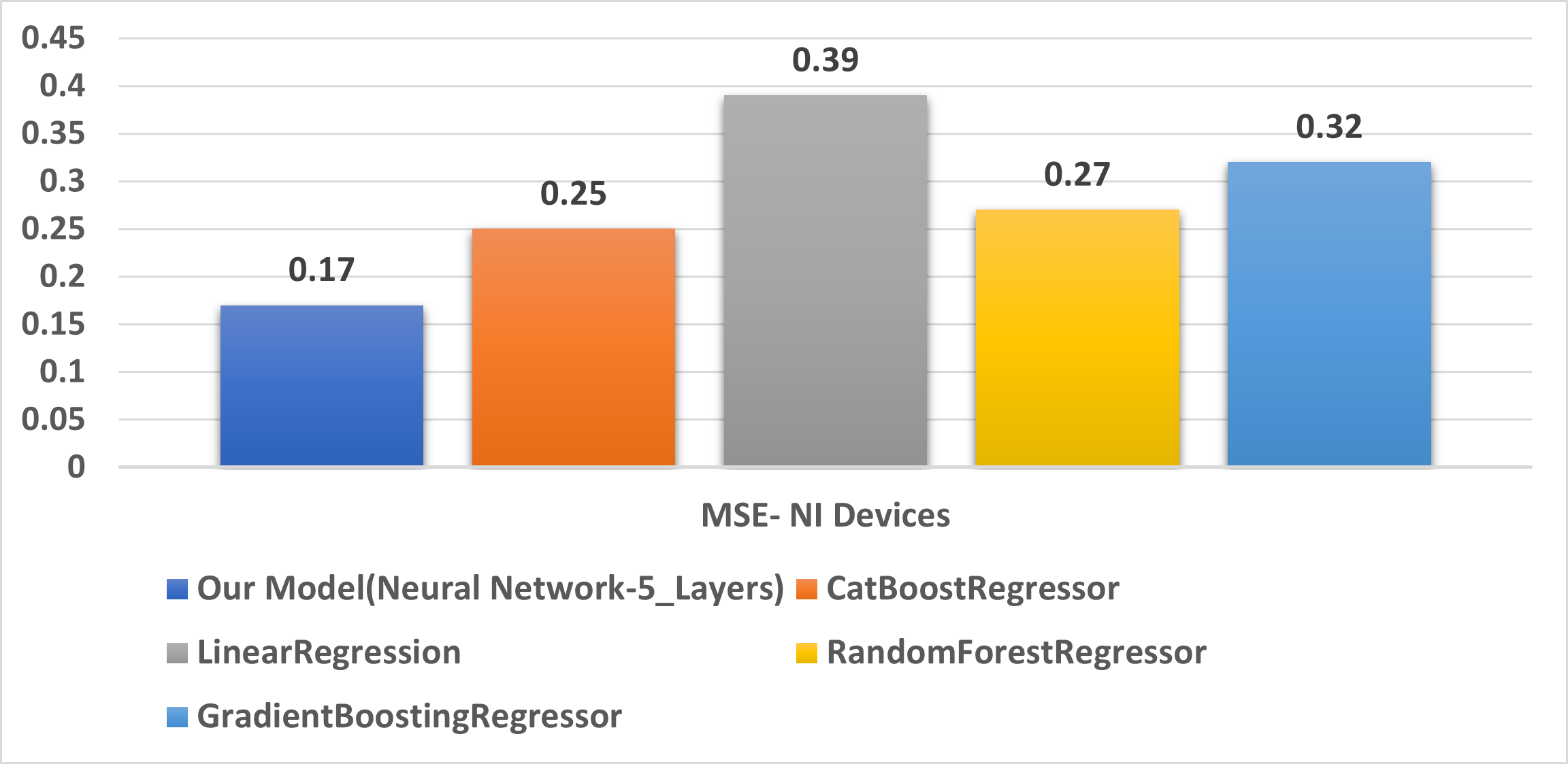}}
        \caption{MSE value for NI dataset trained ML models.}
        \label{fig:MSE-NI}
     \end{subfigure}
    
     \begin{subfigure}[b]{0.45\textwidth}
        \centerline{\includegraphics[scale=0.4]{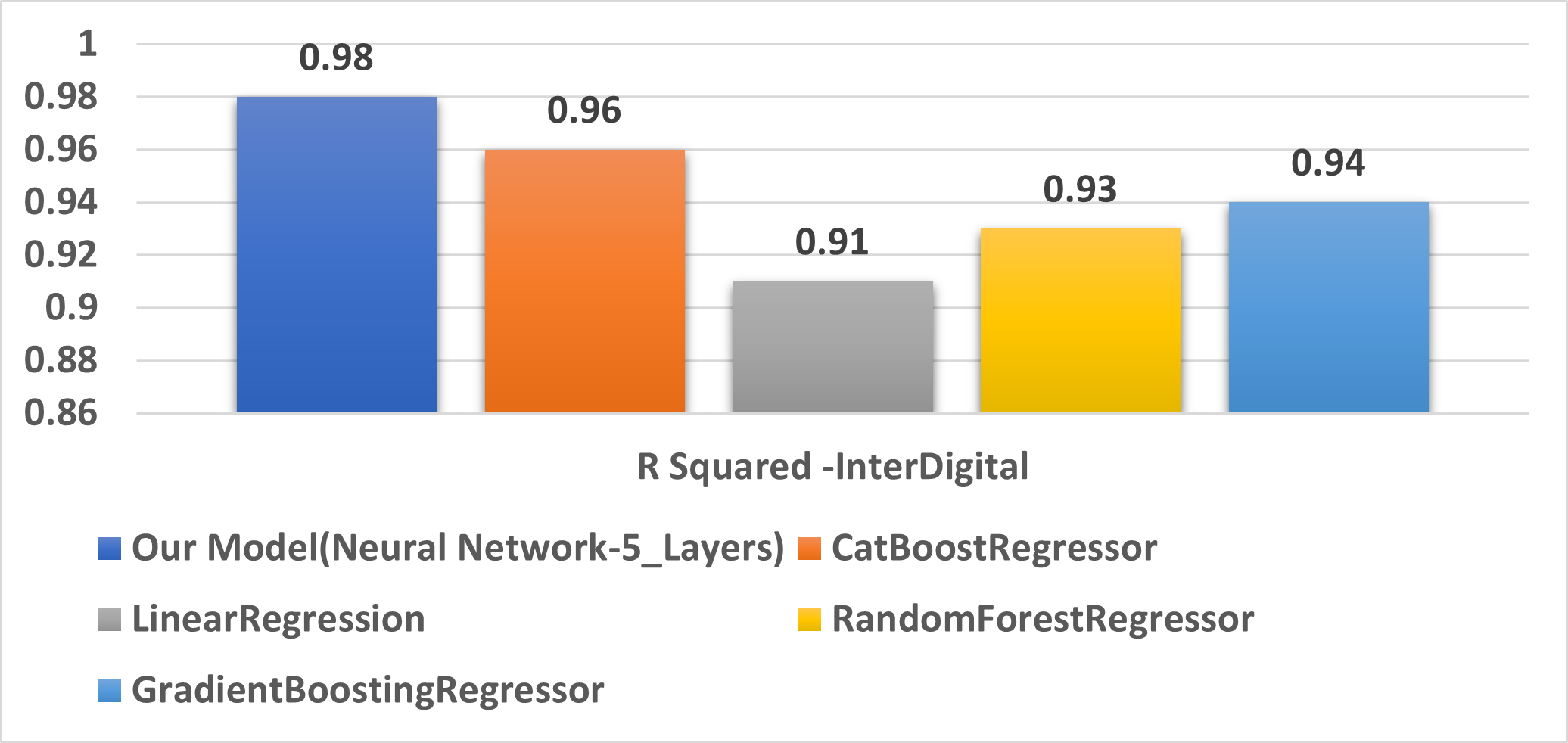}}
        \caption{R2 value for InterDigital dataset trained ML models.}
        \label{fig:ML-Inter}
     \end{subfigure}
     \hfill
     \begin{subfigure}[b]{0.45\textwidth}
        \centerline{\includegraphics[scale=0.4]{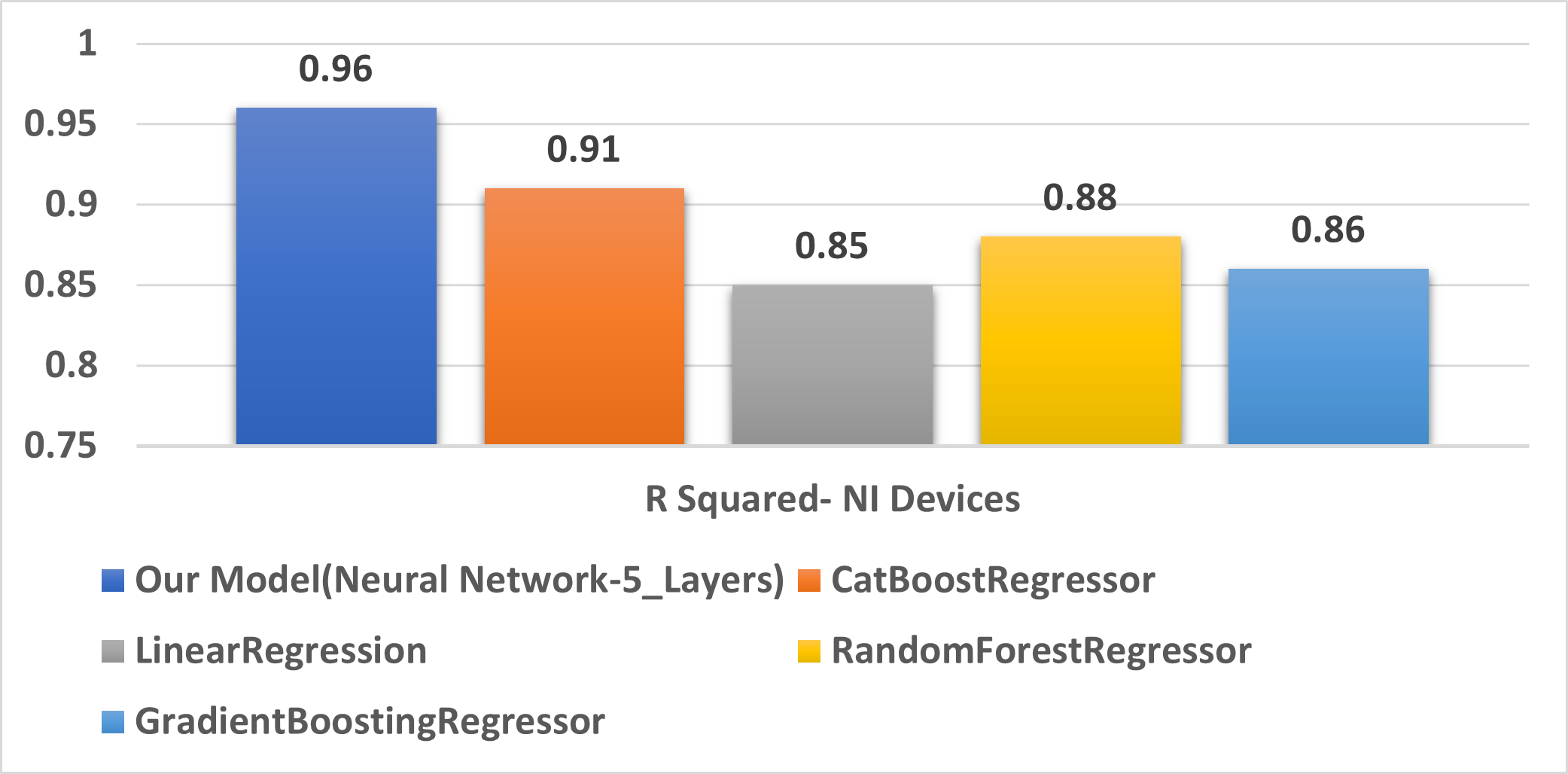}}
        \caption{R2 value for NI dataset trained ML models.}
        \label{fig:ML-NI}
     \end{subfigure}
     \caption{Performance comparison between the designed NN model with other classical ML models. }
     \label{fig:ML-comparison}
\end{figure*}

Given such a comparison, \figurename{ \ref{fig:MSE-Inter}} and \figurename{ \ref{fig:MSE-NI}} show the measured MSE for the selected classical ML models trained on the InterDigital and NI datasets, respectively. For the InterDigital dataset, the NN model with 5 hidden layers achieves the lowest MSE value of 0.16, followed by the CatBoostegressor with the second-lowest value, as shown in \figurename{ \ref{fig:MSE-Inter}}. The Gradient Boosting Regressor achieved relatively higher MSE values compared to the other previously mentioned models. In contrast, Linear Regression stands out as the worst-performing model, obtaining the highest value of the loss function with 0.35. Similarly, for the NI dataset shown in \figurename{ \ref{fig:MSE-NI}}, the designed NN model with 5 hidden layers  obtained the lowest MSE value of 0.17 compared to the other models. CatBoostRegressor and Random Forest Regressor achieved MSE values of 0.25 and 0.27, respectively. As with the InterDigital dataset, Linear Regression performed the worst on the NI dataset as well. Overall, the results from both datasets indicate that the proposed NN model with 5 hidden layers perform the best among the compared ML models, while Linear Regression performs the worst.

On the other hand, the obtained R2 values after training the designed NN models using the measured InterDigital and NI datasets are demonstrated in \figurename{ \ref{fig:ML-comparison}}. Confirming the previous observation, the designed model with 5 hidden layers shows also a significant improvement, with an R2 value of 0.98 and 0.96 for the InterDigital and NI datasets, respectively. Based on the aforementioned evaluation, the results shows that the NN-5layers model outperformed the other prediction models, achieving the lowest MSE and highest R2 values when being trained to predict the RSRP and data rate values for both InterDigital and NI datasets. The CatBoostRegressor also shows good performance in both datasets, while Linear Regression was the worst-performing model.

\subsection{Comparison with the state of the art method}

\begin{figure*}[t]
\centerline{\includegraphics[scale=0.31]{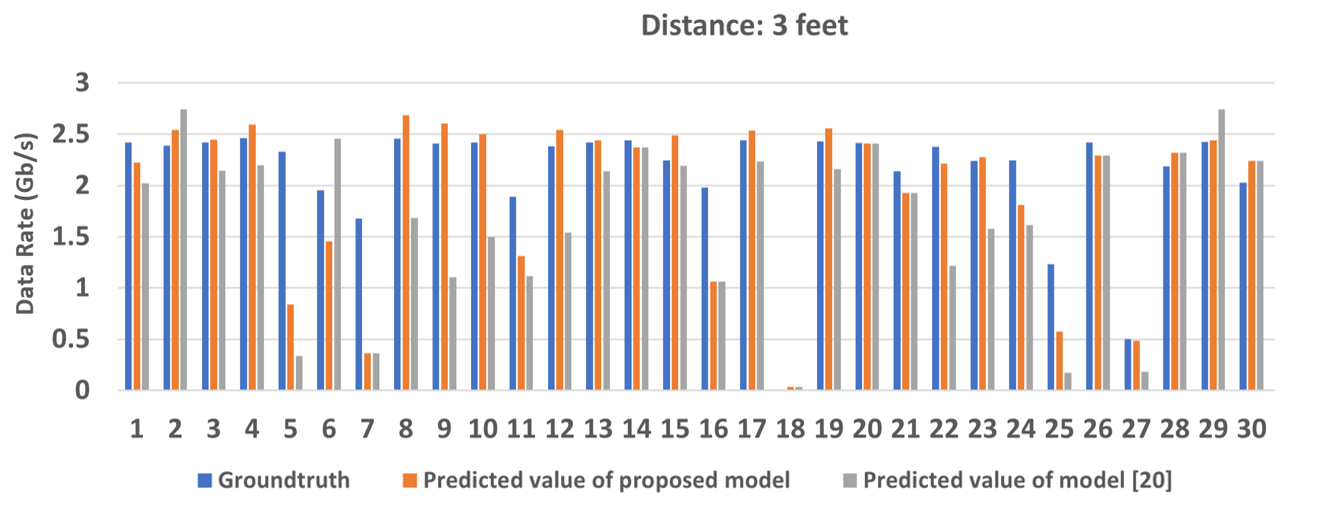}}
\caption{The comparison between the ground truth data rate with the predicted value for our model and~\cite{kazemi2022channel} when TX and RX are placed 3 feet away.  }
\label{fig:diffComparison}
\end{figure*}

In a recent study \cite{kazemi2022channel}, an ML model was proposed for predicting the signal-to-noise ratio (SNR) of a user in a neighboring cell based on the received signal in the serving cell. The algorithm utilizes channel charts and employs an ML model, which is a NN consisting of three layers.  
The input layer of the works in \cite{kazemi2022channel} receives the two-dimensional location of the channel chart and the current beam SNR value, while the output layer provides the predicted SNR values for the other beams. The objective of the algorithm in is to accurately predict the SNR of a user in a neighboring cell, based on the received signal in the serving cell, which can enable efficient resource allocation and enhance the overall network performance. 

We implemented the model developed in \cite{kazemi2022channel} and compared its performance with our designed model in the dataset collected by the NI devices. Due to space constraints, we have selected one scenario as a representative sample for the purpose of illustrating the comparison, i.e., the distance between the TX and RX to 3 feet, and randomly sampled 30 data pairs for comparison.  \figurename{ \ref{fig:diffComparison}}  is indicative of the performance of the two models in predicting the output variable of interest, and serves to highlight the superiority of our proposed approach among most of  tested samples.  We observe that the designed NN model offers a promising solution for predicting the data rate of a TX/RX pairs in various settings. 

%% file: sections/Conclusion.tex
\section{Conclusion and Future Work}
In conclusion, this study conducted an experimental analysis of mmWave beam profiling using two testbeds with different frequency ranges provided by InterDigital (27 GHz) and NI (71 GHz). The investigation aimed to evaluate the communication quality in mmWave configured networks in various reception conditions, where two datasets were generated for this purpose. A neural network model was also developed and trained to predict network transmission quality based on the obtained datasets. Future work will explore mmWave beam profiling under diverse environmental conditions, including static and dynamic obstacles, network requirements, and multiple active TXs. This study's findings contribute to improving mmWave communication systems, and the proposed approach holds potential for future research in this field.

\section*{Acknowledgment}
This work was supported by Commonwealth Cyber Initiative (CCI), an investment in the advancement of cyber R\&D, innovation, and workforce development. Visit CCI at: \url{www.cyberinitiative.org}. In addition, we would like to thank InterDigital for providing us with their mmWave equipment.  This work is supported by the National Science Foundation under Grants CNS-2120279, CNS-1950704, and DUE-1742309, the National Security Agency under Grants H98230-22-1-0275, H98230-21-1-0165, H98230-21-1-0263, and H98230-21-1-0278, the Air Force Research Lab under Grant FA8750-19-3-1000, and InterDigital Communications, Inc.